 \documentclass[final,5p,times,twocolumn]{elsarticle}
 \usepackage{graphicx}
\usepackage{bm}
\usepackage{txfonts}
\usepackage{hyperref}
\usepackage{mathrsfs}
\usepackage[english]{babel}
\usepackage{amsmath}
\usepackage{cleveref}
\usepackage[autostyle, english = american]{csquotes}
\MakeOuterQuote{"}

\date{\today}
\newcommand{\be}{\begin{eqnarray}}
	\newcommand{\ee}{\end{eqnarray}}

\newcommand{\bfz}{{\bf 0}_{\perp}}

\newcommand{\bfk}{{\bf k}_{\perp}}

\newcommand{\bfP}{{\bf P}_{\perp}}

\newcommand{\Dp}{{\bf \Delta}_{\perp}}

\usepackage{xcolor}

\usepackage{amssymb}
\usepackage{lipsum}

\journal{Physics Letters B}

\begin{document}

\begin{frontmatter}




\title{Valence quark properties of charged kaons in symmetric nuclear matter}


	\affiliation{Computational High Energy Physics Lab, Department of Physics, Dr. B.R. Ambedkar National
		Institute of Technology, Jalandhar, Punjab, 144008, India}

\author{Reetanshu Pandey}
\ead{reetanshuhep@gmail.com}
\author{Satyajit Puhan}
\ead{puhansatyajit@gmail.com}
\author{Navpreet Kaur}
\ead{knavpreet.hep@gmail.com}
\author{Arvind Kumar}
\ead{kumara@nitj.ac.in}
\author{Suneel Dutt}
\ead{dutts@nitj.ac.in}
\author{Harleen Dahiya}
\ead{dahiyah@nitj.ac.in}

\begin{abstract}
We calculate the leading twist valence quark transverse momentum parton distribution functions (TMDs) and generalized parton distributions (GPDs) of the charged kaons in an isospin symmetric nuclear matter at zero temperature by employing the light-cone quark model.
The medium modifications of the unpolarized TMDs and GPDs have been carried out by taking inputs from the chiral SU($3$) quark mean field model. The electromagnetic form factors (EMFFs) and charge radii have been calculated from the unpolarized GPDs for both the vacuum and in-medium cases. We have also calculated the variation of average transverse and longitudinal momenta for the active quark at high baryonic density. These results are found to be in good agreement with the available experimental data as well as with other model predictions.
\end{abstract}


\end{frontmatter}

\section{Introduction} 
\label{introduction}
The properties of kaons in nuclear matter have recently garnered significant interest due to their potential to indicate chiral symmetry restoration and provide insights into the possibility of kaon condensation in neutron stars \cite{Tsushima:2000re}. Kaons play a very important and special role in understanding the nonperturbative
features of low-energy quantum chromodynamics (QCD). The nonperturbative aspect of QCD can be studied using quark distributions of these mesons at low $Q^2$, which can be further evolved to perturbative region at high $Q^2$ to access the QCD phenomena experimentally \cite{Goeke:2001tz,Belitsky:2005qn}. Even though the kaons play a key role in understanding electroweak interactions and CP violation, they have received less attention compared to pions \cite{Aguilar:2019teb}. Kaon is an important hadron to study due to the presence of strange quark in it.

This study aims to investigate the in-medium modifications of valence quark distribution function of $K^+$ meson in symmetric nuclear matter using the light-cone quark model (LCQM) \cite{Brodsky:1997de,Acharyya:2024enp,Qian:2008px}, with in-medium inputs derived from the chiral  SU(3) quark mean field model (CQMF) \cite{Wang:2001jw,Kaur:2024wze,Puhan:2024xdq}. The primary constituents: valence quark and antiquark,  which are strongly bound to form the complex hadron structure, play an important role in describing the internal structure of mesons which can be understood using the multidimensional structure of the constituent quarks, gluons and sea-quarks. In this perspective, three dimensional transverse momentum parton distribution functions (TMDs) \cite{Diehl:2015uka,Angeles-Martinez:2015sea,Pasquini:2008ax} and generalized parton distributions (GPDs) \cite{Diehl:2003ny,Chavez:2021llq,Zhang:2021tnr,Broniowski:2022iip,Guidal:2004nd} play a vital role. TMDs being the function of longitudinal momentum fraction ($x$) carried by the constituent quark from parent hadron and transverse momentum of quark ($\textbf{k}_\perp$) carry the information about the orbital motion, intrinsic transverse motion and spin polarization of the parton inside a hadron. However, it lacks the information about the spatial structure of quarks inside the hadron, which can be understood through the study of valence quark GPDs. GPDs encode the information about the physical properties like charge distribution, charge radius, magnetic moment, quadrupole moment, hadron spin, azimuthal asymmetry, and mechanical properties like pressure distribution, force distribution, mechanical radius, shearing stress, energy-momentum tensor etc. GPDs are  a function of longitudinal momentum fraction ($x$), skewness parameter $\xi$ (controls the off-forwardness of the process) and momentum transfer between initial and final hadron ($\Delta$). GPDs can be extracted from deeply virtual Compton scattering process (DVCS) \cite{Ji:1996nm} and deeply virtual meson production (DVMP) \cite{Favart:2015umi} process experimentally. While TMDs can be accessed through deep inelastic scattering, semi-inclusive deep inelastic scattering (SIDIS) \cite{Bacchetta:2017gcc}, Drell-yan process \cite{Zhou:2009jm} and $Z^0/W^\pm$ process experimentally. Both TMDs and GPDs give rise to the one dimensional parton distribution functions (PDFs) with proper assumptions. PDFs show the probability of finding the constituent quark inside a hadron as a function of longitudinal momentum fraction ($x$) \cite{Martin:1998sq}. 

In this work, we have calculated the leading twist time reversal (T-even) TMDs and chiral-even GPDs for the case of kaons in LCQM in free space and further extended to study the effect in symmetric nuclear medium at different baryonic densities using the CQMF model. LCQM is a non-perturbative approach for studying the internal structure and properties of hadrons, such as mass spectra, radiative decays, and decay constants \cite{Brodsky:1997de,Acharyya:2024enp,Qian:2008px,Puhan:2024jaw}. It is inherently relativistic and gauge-invariant. One of its main strengths is its focus on valence quarks, which are the key contributors to the hadron’s overall structure and characteristics. Even after evolving through QCD corrections into the perturbative regime, LCQM remains highly effective in describing hadronic structure in the form of EMFFs, PDFs, TMDs, GPDs etc. While CQMF model confines the quarks inside the baryons through confining potential and interact internally via different fields.

The unpolarized quark GPDs of kaon at the leading twist carry information about the electromagnetic form factors (EMFFs) and charge radii of both quark and hadrons along with the PDFs, whereas the quark TMDs provide a benchmark to the transverse structure of the quark inside it. It is an important and interesting topic to understand how the surrounding nuclear medium affect the behavior and properties of valence quarks of kaon. The effect of medium modifications on nucleon structure function $F^N_2$ in the Bjorken range of $0.2<x_B<0.7$ was first verified by European Muon Collaboration (EMC) \cite{EuropeanMuon:1983wih} and later in the DIS scattering cross section of different nuclei by Stanford Linear Accelerator Center (SLAC) \cite{Arnold:1983mw}. These results indicated that the baryonic density has considerable effect on the internal structure of hadron, mostly on the valence quark and antiquark. This EMC effect is caused due to the strong short-range correlations  functions (SRC) among the nucleons. The vacuum leading twist GPDs and TMDs have been calculated using the quark-quark correlator by taking the minimal Fock-state of kaon i.e, $|\Psi_K\rangle=\Psi_{u \bar s}|q \bar q\rangle$. There have been many predictions to study the internal structure of pions through leading twist quark TMDs and GPDs in different models in vacuum as well as in the nuclear medium \cite{Kaur:2024wze,Puhan:2024xdq,Chavez:2021llq,Kaur:2018ewq,Son:2024uet,Arifi:2024tix,Hutauruk:2023ccw}, while very few studies have been predicted for kaons \cite{Yabusaki:2023zin}. Kaon medium modifications have been studied in the kaon photo production off nuclei \cite{Lee:1999kd} and kaon-nucleus Drell-Yan processes \cite{Londergan:1996vh}.

This paper is arranged as follows: In Sec. \ref{SecModel}, we present the details of LCQM and CQMF.
The unpolarized TMDs and GPDs are calculated in Sec. \ref{sectmd} and \ref{secgpd}, respectively. In Sec. \ref{secres}, the results of the present work are discussed, whereas summary and conclusion is given in Sec. \ref{secsum}.

\section{The models}
\label{SecModel}
\subsection{Light-cone quark model}
The leading order light-cone (LC) Fock-state meson wave function in terms of quark-antiquark pair can be expressed as \cite{Qian:2008px,Pasquini:2023aaf,Lepage:1980fj}
\begin{eqnarray}
    |\Psi_{K}(P^+,\bfP)\rangle=\sum_{\lambda_q,\lambda_{\bar q}} \int  \frac{dx d^2\bfk}{2(2\pi)^3\sqrt{x(1-x)}}
 \psi^{\lambda_q, \lambda_{\bar q}}_{K}(x,\bfk)
 \nonumber\\
 |q (xP^+, \bfk,\lambda_q) {\bar q}((1-x)P^
 +, -\bfk, \lambda_{\bar q})\rangle \, .
\label{MesonState}
\end{eqnarray}
Here, $\Psi_{K}$ is the kaon eigen state with four-vector momentum $P=(P^+,P^-,\bfP)$. $\lambda_{q(\bar q)}$ and $k=(k^+,k^-,\bfk)$ are the quark (antiquark) helicity and four-vector momentum of quark respectively. $x=k^+/P^+$ is the longitudinal momentum fraction carried by the quark from its parent hadron and simultaneously $1-x$ is the fraction carried by the antiquark. For this work, we mainly focus on the valence quark distributions by not considering the gluon contribution. 
\par The four-vector momenta of kaon, quark and antiquark respectively are expressed in the light-cone frame as

\be 
P &=& \bigg(P^+,\frac{M_{K}^{\ast2}}{P^+},\bfz\bigg) \, ,  \nonumber \\
k_1 &=& \bigg(x P^+,\frac{\bfk^2 + m^{\ast2}_u}{x P^+},\bfk \bigg) \, ,  \nonumber \\
k_2 &=& \bigg((1-x) P^+,\frac{\bfk^2 + m^{\ast2}_{\bar{s}}}{(1-x) P^+},-\bfk \bigg) \, . 
\ee
Here, $m^*_u$ and $m^*_{\bar{s}}$ are the  effective masses for $u$ and $\bar{s}$ quarks, respectively. $M_K^*$  is the in-medium  bound state mass of kaon expressed as 
\be
M_{K}^* = \sqrt{\frac{m_u^{*2}+\bfk^2}{x}+\frac{m_{\bar s}^{*2}+\bfk^2}{(1-x)}}.
\ee
The term $\psi^{\lambda_q, \lambda_{\bar q}}_{u \bar s}(x,\bfk)$ in Eq. (\ref{MesonState}) is the spin improved total wave function of kaon  defined as \cite{Huang:1994dy} 
\begin{eqnarray}
    \psi^{\lambda_q, \lambda_{\bar q}}_{K}(x,\bfk)= \varphi_{K}(x,\bfk) \, \chi^{\lambda_q, \lambda_{\bar q}}_K (x,\bfk) \, ,\label{sm}
\end{eqnarray}
where $\varphi_{K}(x,\bfk)$ and $\chi^{\lambda_q, \lambda_{\bar q}}_K (x,\bfk)$ are the momentum space wave function and spin wave function of kaon, respectively. For this work, we have adopted Brodsky-Huang-Lepage prescription, which can be expressed as 
\cite{Kaur:2020vkq,Xiao:2002iv,Yu:2007hp} 

\be 
\varphi_{K} (x,\bfk) &=& A_K \, exp \,\Biggl[-\frac{M_K^{\ast2}}{8 \beta_K^2} - \frac{(m_u^{\ast2} - m_{\bar{s}}^{\ast2})^2}{8 \beta_K^2 M_K^{\ast2}}\Biggr] \, ,
\ee 
where $A_K$ and $\beta_K$ are the normalization constant and harmonic scale parameter which have been fixed with the decay constant of kaon \cite{Kaur:2019jow}. 
\par The normalization constant $A_K$ can be calculated using
\begin{eqnarray}
    \int \frac{dx d^2\bfk}{16 \pi^3} |\varphi_{K} (x,\bfk)|^2=1.
\end{eqnarray}
The spin wave function $\chi^{\lambda_q, \lambda_{\bar q}}_K (x,\bfk)$ in light-cone dynamics can be calculated by two different methods, either from Melosh-Wigner rotation or from quark-meson vertex. Both the methods give rise to same results. In this work, we have calculated the spin wave function from the quark-meson vertex by choosing proper gamma matrices and Dirac matrices as \cite{Qian:2008px,Choi:1996mq}
\begin{eqnarray}
    \chi_K^{\lambda_q,\lambda_{\bar q}}(x,\textbf{k}_\perp) = \bar u (k_1,\lambda_q) \Gamma_m  v(k_2,\lambda_{\bar q}) \, ,
\end{eqnarray}
with $\Gamma_m= \frac{\gamma_5}{\sqrt{2}\sqrt{M_K^{*2}-(m_{u}^{*}-m_{\bar s}^{*})^{2}}}$. $u$ and $v$ are four-vector Dirac matrices. Using different helicities of quark and antiquark, the spin wave functions are found to be 

\begin{equation}
  \begin{array}{lll}
    \chi_K^{\uparrow, \uparrow}(x,\mathbf{k}_\perp)&=&\frac{1}{\sqrt{2}}\omega^{-1}(-\textbf{k}_1+\iota \textbf{k}_2),\\
    \chi_K^{\uparrow, \downarrow}(x,\mathbf{k}_\perp)&=&\frac{1}{\sqrt{2}}\omega^{-1}((1-x)m^*_u+x m^*_{\bar s}),\\
    \chi_K^{\downarrow, \uparrow}(x,\mathbf{k}_\perp)&=&\frac{1}{\sqrt{2}}\omega^{-1}(-(1-x)m^*_u-x m^*_{\bar s}),\\
    \chi_K^{\downarrow, \downarrow}(x,\mathbf{k}_\perp)&=&\frac{1}{\sqrt{2}}\omega^{-1}(-\textbf{k}_1-\iota \textbf{k}_2) .
  \end{array}
\end{equation}
Here, $\omega= \sqrt{x(1-x)M_K^{*2}-(m_{u}^{*}-m_{\bar s}^{*})^{2}}$. Similarly, the spin wave function can also be calculated by transforming instant form to front form through Melosh-Wigner rotation which will provide the same result. 
\par The meson state wave function in Eq. (\ref{MesonState}) with different quark-antiquark helicity can be written explicitly as 

\be 
\Psi_{K}(P^+,\bfP)\rangle &=& \int \frac{dx \, d^2 \bfk}{  2 (2\pi)^3 \sqrt{x(1-x)}} \, \nonumber \\ &\times& \big[ \psi^{\uparrow, \uparrow}_{K}(x,\bfk)
 |q (1,\uparrow) {\bar q}(2, \uparrow)\rangle \,    \nonumber \\
&+& \psi^{\uparrow, \downarrow}_{K}(x,\bfk)
 |q (1,\uparrow) {\bar q}(2 \downarrow)\rangle  \nonumber \\
&+& \psi^{\downarrow, \uparrow}_{K}(x,\bfk)
 |q (1,\downarrow) {\bar q}(2, \uparrow)\rangle \nonumber \\ &+& \psi^{\downarrow, \downarrow}_{K}(x,\bfk)
 |q (1,\downarrow) {\bar q}(2, \downarrow)\rangle \big] \, .\label{eqnq}
\ee 
Here, we have denoted $1$ and $2$ in place of $(x P^+, \bfk)$ and $((1-x)P^+,-\bfk)$ for quark and antiquark momentum, respectively.
For numerical calculations, quark masses $(m^*_{u(\bar s)})$ and harmonic scale parameter $\beta_K$ are required. For vacuum case, these parameters have been taken from Ref. \cite{Puhan:2024jaw}, which are then modified for the study of in-medium properties in CQMF.

\subsection{Chiral SU(3) quark mean field model (CQMF)}\label{cqmf}

The CQMF model provides a framework for calculating the effective masses of quarks in the nuclear medium, drawing from the low-energy dynamics of QCD \cite{Wang:2001jw} and the concept of broken scale invariance. This model describes the  interactions between quarks and mesons as well as the meson-meson interactions, across a wide range of temperatures and densities. The quark-meson interactions arise from the exchange of scalar ($\sigma$, $\zeta$, and $\delta$) and vector ($\omega$ and $\rho$) fields, leading to modifications in baryon properties based on their constituent quark flavors in an asymmetric nuclear medium.
The non-strange scalar-isoscalar field $\sigma$ is linked to the $f_0$ meson ($\sim 500$ MeV), composed of light quark-antiquark pairs ($u, d, \bar{u}, \bar{d}$). Meanwhile, the $\zeta$ field corresponds to a scalar meson with strange quark content, incorporating the effects of strangeness in the medium. To account for isospin asymmetry, the model includes the scalar isovector field $\delta$ and the vector isovector field $\rho$. Additionally, the dilaton field $\chi$ plays a crucial role in implementing the broken scale invariance of QCD \cite{Wang:2001jw}.
The thermodynamic potential for isospin-asymmetric nuclear matter at finite temperature and density is given by
\begin{equation}
\begin{split}
\Omega = -\frac{k_B T}{(2\pi)^3} \sum_i \gamma_i \int_{0}^{\infty} d^3k \Bigg\{ 
\ln\left(1+e^{-(E_i^*(k)-\nu_i^*)/k_B T}\right) \\
+ \ln\left(1+e^{-(E_i^*(k)+\nu_i^*)/k_B T}\right) \Bigg\} - \mathcal{L}_M - V_{\text{vac}}.
\end{split}
\label{thermo}
\end{equation}
The summation runs over nucleons in the medium, represented by \( i = p/n \), with a degeneracy factor \( \gamma_i = 2 \) accounting for the two spin states of each nucleon. The effective energy of a baryon \( E_i^*(k) \) is given by

\[
E_i^*(k) = \sqrt{M_i^{*2} + k^2},
\]
where \( M_i^* \) is the effective baryon mass. To ensure zero vacuum energy,  \( V_{vac} \) is subtracted. The effective chemical potential of a baryon \( \nu_i^* \) is related to the free chemical potential \( \nu_i \) as
\begin{equation}
\nu_i^* = \nu_i - g_{\omega }^{i} \omega - g_{\rho }^{i} I^{3i} \rho,
\end{equation}
where $g_{\omega }^{i}$, and $g_{\rho }^{i}$ are the coupling constants of the vector fields $\omega$ and $\rho$, respectively. The term, $\mathcal{L}_M$, in Eq. (\ref{thermo}) is mesonic Lagrangian, which can be written as \cite{Wang:2001hw}
\begin{equation}
\mathcal{L}_M = \mathcal{L}_{\Sigma \Sigma} + \mathcal{L}_{VV} + \mathcal{L}_{SB}.
\end{equation}
Here, $\mathcal{L}_{\Sigma \Sigma}$ represents the self-interaction of scalar mesons and, in the mean-field approximation, is expressed as \cite{Heide:1993yz}
\begin{align}
\mathcal{L}_{\Sigma \Sigma} &= -\frac{1}{2}k_0 \chi^2 (\sigma^2 + \zeta^2 + \delta^2) + k_1 (\sigma^2 + \zeta^2 + \delta^2)^2 \nonumber \\
&+ k_2 \left(\frac{\sigma^4}{2} + \frac{\delta^4}{2} + 3\sigma^2\delta^2 + \zeta^4\right) + k_3 \chi (\sigma^2 - \delta^2) \zeta \nonumber \\
&- k_4 \chi^4 - \frac{1}{4}\chi^4 \ln\frac{\chi^4}{\chi_0^4} + \frac{\xi}{3}\chi^4 \ln\left(   \left( \frac{(\sigma^2 - \delta^2) \zeta}{\sigma_0^2 \zeta_0} \right) \left( \frac{\chi^{3}}{\chi^{3}_{0}}\right) \right).
\label{lagrang}
\end{align}
The scale-breaking effect is introduced by the last two logarithmic terms, which are used to compute the trace of energy-momentum tensor within this model. The term for vector meson self-interactions \( \mathcal{L}_{VV} \) is represented as
\begin{equation}
\mathcal{L}_{VV} = \frac{1}{2}\frac{\chi^2}{\chi_0^2} \left(m_\omega^2 \omega^2 + m_\rho^2 \rho^2\right) + g_4 (\omega^4 + 6\omega^2 \rho^2 + \rho^4).
\end{equation}
The Lagrangian density term $\mathcal{L}_{SB}$ is responsible for the explicit breaking of chiral symmetry, which generates non-zero masses for pseudoscalar mesons and can be expressed as \cite{Papazoglou:1998vr}

\begin{equation}
\mathcal{L}_{SB} = \frac{\chi^2}{\chi_0^2} \left[m_\pi^2 \kappa_\pi \sigma + \left(\sqrt{2}m_K^2 \kappa_K -\frac{m^{2}_{\pi}}{\sqrt{2}} \kappa_{\pi}\right) \zeta \right].
\end{equation}
The CQMF model incorporates a confinement mechanism for quarks within baryons via a Lagrangian term 

\[
\mathcal{L}_c = - \bar{\Psi} \chi_c \Psi.
\]
In the presence of meson mean fields, the Dirac equation governing a quark field \( \Psi_{qi} \) is expressed as

\begin{equation}
\left[-i {\vec{\alpha}} \cdot \vec{\nabla} + \chi_c(r) + \beta m_q^*\right] \Psi_{qi} = e_q^* \Psi_{qi}.
\end{equation}
Here, the subscripts \( q \) and \( i \) represent the quark \( q \) (where \( q = u, d, s \)) within a baryon of type \( i \) (where \( i = p, n \)). The effective quark mass \( m_q^* \) and energy \( e_q^* \) are described by the following relations in terms of scalar fields (\( \sigma \), \( \zeta \), and \( \delta \)) and vector fields (\( \omega \) and \( \rho \)) as

\begin{equation}
m_q^* = -g_\sigma^q \sigma - g_\zeta^q \zeta - g_\delta^q I^{q}_{3} \delta + \Delta m_q,
\end{equation}
and
\begin{equation}
e_q^* = e_q - g_\omega^q \omega - g_\rho^q I^{q}_{3} \rho,
\end{equation}
respectively. The coupling constants of quarks with the scalar fields \( \sigma \), \( \zeta \), and \( \delta \) are represented by \( g_\sigma^q \), \( g_\zeta^q \), and \( g_\delta^q \), respectively. The third component of isospin for the quark flavor \( I_3^q \) is given by $I_3^{u} = \frac{1}{2}, \quad I_3^{d} = -\frac{1}{2}, \quad I_3^{s} = 0.$ The additional mass terms $\Delta m_{u/d} = 0 \quad \text{and} \quad \Delta m_s = 77 \, \text{MeV}$ are determined to fit the vacuum masses of the quarks. The effective mass of the baryon, \( M_i^* \), is related to the spurious center of momentum \( \langle p_{i\, \text{cm}}^2 \rangle \) and the effective quark energy \( e_q^* \) as \cite{Wang:2001hw}
\begin{equation}
M_i^* = \sqrt{\left(\sum_q n_{qi} e_q^* + E_{i\, \text{spin}}\right)^2 - \langle p_{i\, \text{cm}}^2 \rangle}.
\end{equation}
Here, \( n_{qi} \) represents the number of \( q \)-flavored quarks in the \( i^{\text{th}} \) baryon. The term \( E_{i\, \text{spin}} \) serves as a correction factor to the baryon energy arising from spin-spin interactions and is calibrated to reproduce the baryon vacuum mass. The spurious center of momentum of the baryon, \( \langle p_{i\, \text{cm}}^2 \rangle \), can be expressed in terms of \( e_q^* \) and \( m_q^* \) via the following relation \cite{Barik:2013lna}
\begin{equation}
\langle p_{i\, \text{cm}}^2 \rangle = \sum_q \frac{(11 e_q^* + m_q^*)}{6(3e_q^* + m_q^*)} \left( e_q^{*2} - m_q^{*2} \right).
\end{equation}
The total thermodynamic potential, defined in Eq. (\ref{thermo}), is minimized with respect to the mesonic fields \( \phi \), where \( \frac{\partial \Omega}{\partial \phi} = 0 \), to compute the coupled equations of motion for these fields. These equations are as follows
\begin{align}
k_0 \chi^2 \sigma - 4 k_1 (\sigma^2 + \zeta^2 + \delta^2) \sigma - 2k_2 (\sigma^3 + 3\sigma\delta^2) - 2k_3 \chi  \sigma \zeta  \notag \\
- \frac{\xi}{3} \chi^{4} \left(\frac{2 \sigma}{\sigma^2 - \delta^2}\right) 
+ \left(\frac{\chi}{\chi_0}\right)^2 m^{2}_{\pi }\kappa_\pi 
+ \left(\frac{\chi}{\chi_0}\right)^2 m_{\omega}\omega^2 \frac{\partial m_\omega}{\partial\sigma} \notag \\
- \left(\frac{\chi}{\chi_0}\right)^2 m_{\rho} \rho^2\frac{\partial m_\rho}{\partial \sigma} 
= \sum_{i=p,n} g_{\sigma i} \rho_{i}^{s}.
\label{set12}
\end{align}
\begin{align}
k_0 \chi^2 \zeta - 4 k_1 (\sigma^2 + \zeta^2 + \delta^2) \zeta - 4k_2 \zeta^3 - k_3 \chi (\sigma^2 - \delta^2) - \frac{\xi}{3} \frac{\chi^{4}}{\zeta} \left(\frac{\zeta}{3}\right) \nonumber \\
+ \left(\frac{\chi}{\chi_0}\right)^2 \left[ \sqrt{2}m_k^2 \kappa_K - \frac{m^{2}_{\pi}\kappa_{\pi}}{\sqrt{2}} \right] = \sum_{i=p,n} g_{\zeta i} \rho_{i}^{s},
\end{align}
\begin{align}
k_0 \chi^2 \delta - 4 k_1 (\sigma^2 + \zeta^2 + \delta^2) \delta - 2k_2 (3\delta\sigma^2 + \delta^3) - 2 k_3 \chi \delta \zeta \nonumber \\
- \frac{\xi}{3} \chi^{4} \left(\frac{2 \delta}{\sigma^2 - \delta^2}\right) = \sum_{i=p,n} g_{\delta i} I^{i}_{3} \rho_{i}^{s},
\end{align}
\begin{align}
k_0 \chi \left(\sigma^2 + \zeta^2 + \delta^2 \right) 
- k_3 \chi \left(\sigma^2 - \delta^2 \right) \zeta \notag \\
+ \Bigg[4 k_1 + 1 - \ln\frac{\chi^4}{\chi_0^4} 
- \frac{4d}{3} \ln\left(\frac{(\sigma^2 - \delta^2)\zeta}{\sigma_0^2 \zeta_0}\right) \Bigg] \chi^3 \notag \\
+ \frac{2 \chi}{\chi_0^2} 
\Bigg[ m_\pi^2 \kappa_\pi \sigma + \left(\sqrt{2}m_K^2 \kappa_K 
- \frac{m_{\pi}^{2} \kappa_{\pi}}{\sqrt{2}}\right) \zeta \Bigg] 
- \frac{\chi}{\chi_0^2} \left(m_\omega^2 \omega^2 + m_\rho^2 \rho^2\right) = 0.
\end{align}
\begin{align}
\frac{\chi^2}{\chi_0^2} \left(m_\omega^2 \omega + 4 g_4 \omega^3 + 12 g_4 \omega \rho^2\right) = \sum_{i=p,n} g_{\omega i} \rho^{\nu}_i,
\end{align}
\vspace{0.4cm}
\begin{align}
\frac{\chi^2}{\chi_0^2} \left(m_\rho^2 \rho + 4 g_4 \rho^3 + 12 g_4 \omega^2 \rho\right) = \sum_{i=p,n} g_{\rho i} I^{i}_{3} \rho^{\nu}_i.
\label{set17}
\end{align}
Here, \( I_3^{p} = -I_3^{n} = \frac{1}{2} \) represent the third component of isospin for the nucleon. The symbols \( m_\pi \), \( m_K \), \( m_\omega \), and \( m_\rho \) represent the masses of the \( \pi \), \( K \), \( \omega \), and \( \rho \) mesons, respectively. Additionally, \( \rho_i^{\nu} \) and \( \rho_i^{s} \) denote the vector and scalar densities of the \( i^{\text{th}} \) baryon, respectively, and can be expressed as
{\footnotesize \begin{align}
 \rho_{i}^{\nu} &= \gamma_i \int \frac{d^3k}{(2\pi)^3} 
\left[\frac{1}{1 + \exp\left(\beta[E_i^*(k) - \nu_i^*]\right)} 
- \frac{1}{1 + \exp\left(\beta[E_i^*(k) + \nu_i^*]\right)}\right],
\end{align}}
and
{\footnotesize \begin{align}
 \rho_i^{s} &= \gamma_i \int \frac{d^3k}{(2\pi)^3} \frac{m^{*}_{i}}{E^{*}_{i}(k)}  
\left[\frac{1}{1 + \exp\left(\beta[E_i^*(k) - \nu_i^*]\right)} 
+ \frac{1}{1 + \exp\left(\beta[E_i^*(k) + \nu_i^*]\right)}\right].
\end{align}}
The set of non-linear Eqs. (\ref{set12})-(\ref{set17}) are solved for different values of baryon density, temperature, and isospin asymmetry of the medium. The parameter \( \eta \) is defined as

\[
\eta = -\frac{\sum_{i} I_3^{i} \rho_i^{\nu}}{\rho_B},
\]
where \( \rho_B = \rho_p + \rho_n \) represents the total baryonic density of the nuclear medium. The impact of asymmetric parameters $\eta$ and temperature $T$ was found to be less on the distributions  \cite{Singh:2024lra}.
Hence, in the present work, we limit ourselves to symmetric nuclear matter at zero temperature, focusing on the impact of finite density of nuclear matter on the TMDs and GPDs of kaons.

\section{Transverse momentum parton distribution functions (TMDs)}\label{sectmd}

For the case of spin-$0$ pseudoscalar mesons, there are two TMDs present at the leading twist, which are unpolarized $f^q_1(x,\bfk^2)$ and Boer-Mulders function $h^\perp_1(x,\bfk^2)$ quark TMDs. For this work, we have only considered $f^q_1(x,\textbf{k}^2_\perp)$ T-even TMD, which is expressed in terms of quark-quark correlator as \cite{Puhan:2024jaw,Meissner:2008ay}
 \begin{align}
f^q_1(x,\textbf{k}^2_\perp)&=&\frac{1}{2}\int\frac{\mathrm{d}z^-\mathrm{d}^2\vec{z}_\perp}{2(2\pi)^3}e^{i \textbf{k}\cdot z} \langle \Psi_{K}(P^+,\bfP)|\bar\psi(0)\gamma^+ \nonumber \\
    && \times W(0,z)\psi(z)|\Psi_{K}(P^+,\bfP)\rangle|_{z^+=0} ,
    \label{tmd}
\end{align}
where $z=(z^+,z^-,z_\perp)$ is the position four-vector. $\psi(z)$ is the quark field operator and $W(0,z)$ is the Wilson line, which preserves the gauge invariance of the bilocal quark field operators in the correlation
functions. $W(0,z)$ decides the path of the quark field, which is taken to be unity in this case. By substituting kaon eigenstate  in Eq. (\ref{tmd}), the overlap form of valence quark TMD with different helicities of quark and antiquark is found to be 
\begin{eqnarray}
    f^q_1(x,\textbf{k}^2_\perp)&=&\frac{1}{16 \pi^3} \Big(|\psi^{\uparrow,\uparrow}_K (x,\bfk^2)|^2+|\psi^{\uparrow,\downarrow}_K (x,\bfk^2)|^2 \nonumber\\&& +|\psi^{\downarrow,\uparrow}_K (x,\bfk^2)|^2+|\psi^{\downarrow,\downarrow}_K (x,\bfk^2)|^2 \Big).
\end{eqnarray}
The implicit form of $f^q_1(x,\bfk^2)$ is found to be 
\begin{equation}
    f^q_1(x,\bfk^2)=\frac{1}{16 \pi^3} \Big(\bfk^2+((1-x^*)m^*_u+x^* m^*_{\bar s})^2\Big) \frac{|\varphi_{K}(x,\bfk)|^2}{\omega^2}.
\end{equation}
Here, $x^*$ is the in-medium longitudinal momentum fraction, which is connected with vacuum longitudinal momentum fraction as \cite{Puhan:2024xdq}
\begin{align}
 x_j^*  = 
 \begin{cases}
 \frac{E_j^* + g_{\omega}^{j}\omega + 
 g_{\rho}^{j} I^{3j}\rho + k_j^{*3}}{E_j^* + E_{\bar j}^* + g_{\rho}^{{ j}} \left(I^{3{j}}-I^{3{\bar j}}\right)\rho + P^{*3}} = \frac{x_j+ (g_{\omega}^{j}\omega + 
 g_{\rho}^{j} I^{3j}\rho)/P^+}{1+\left(I^{3{j}}-I^{3{\bar j}}\right)\rho/P^+} \quad \text{for quark } q \\
  \frac{E_{\bar j}^* - g_{\omega}^{j}\omega - 
  g_{\rho}^{j} I^{3\bar j}\rho + k_{\bar j}^{*3}}{E_j^* + E_{\bar j}^* + g_{\rho}^{{ j}} \left(I^{3{j}}-I^{3{\bar j}}\right)\rho + P^{*3}} = \frac{x_j - (g_{\omega}^{j}\omega + 
 g_{\rho}^{j} I^{3j}\rho)/P^+}{1+\left(I^{3{j}}-I^{3{\bar j}}\right)\rho/P^+} \quad
  \text{for antiquark } \bar{q} ,
 \end{cases}
 \label{Eq_xfrac_med2}
 \end{align}
In this work, we restrict our calculations to $x$ only. The antiquark TMD can be calculated from the valence quark TMD by the relation
\begin{eqnarray}
    f^q_1(x,\bfk)=f^{\bar q}_1(1-x,-\bfk).
\end{eqnarray}
Both the TMDs should obey the conservation of momentum i.e., the total momentum fraction should be unity and the transverse momenta to be zero.
The average transverse momentum and longitudinal momenta carried by the valence quark can be calculated as 
\begin{eqnarray}
    \langle \bfk^q \rangle = \frac{\int dx d^2\bfk \bfk f^q_1(x,\bfk^2)}{\int dx d^2\bfk  f^q_1(x,\bfk^2)}, \\
    \langle x^{u (\bar s)} \rangle = \frac{\int dx d^2\bfk x f^q_1(x,\bfk^2)}{\int dx d^2\bfk  f^q_1(x,\bfk^2)}.
\end{eqnarray}

\section{Generalized parton distributions (GPDs)}\label{secgpd}
The spatial structure of kaons can be understood through the GPDs. There are total two leading twist quark GPDs for pseudoscalar mesons, out of which $H^q(x,\xi,-\Delta^2_\perp)$ is chiral-even. For this work, we have calculated the GPDs for $\xi=0$. The chiral-even unpolarized $H^q(x,\xi,-\Delta^2_\perp)$ GPDs can be  defined in terms of the bilocal current of quark quark correlator as \cite{Diehl:2003ny,Meissner:2008ay}
\begin{eqnarray}
    H^q(x, 0, -\Delta^2_\perp) &=& \frac{1}{2} \int \frac{dz^-}{2\pi} e^{ix\bar{P}^+z^-} \bigg\langle \Psi_{K}(P^+,\bfP^{\prime\prime})|\bar\psi(0)\gamma^+ \nonumber \\
    &\times& W(0,z)\psi(z)|\Psi_{K}(P^+,\bfP^{\prime})\rangle . 
    \label{gpd}
\end{eqnarray}
 $P^{\prime \prime}$ and $P^\prime$ are the final and initial state meson momentum. Other kinematic variables which include the four-momentum transfer and skewness parameter are respectively expressed as $\Delta^\mu=P^{\prime \prime}-P^\prime$ with $t=\Delta^2=-\Delta^2_\perp$ and $\xi=-\Delta^+/2P^+$. Now utilizing kaon wave function from Eq. (\ref{MesonState}) in the above equation, the valence quark GPD $H^q(x,0,t)$ in overlap form of light-cone wave functions (LCWFs) is found to be 
\begin{eqnarray}
    H^q(x,0,t)=\int \frac{d^2 \mathbf{k_\perp}}{16 \pi^3} \big[\psi_K^{\uparrow,\uparrow\ast}(x,\mathbf{k}^{\prime\prime}_\perp)  \psi_K^{\uparrow,\uparrow} (x,\mathbf{k}^{\prime}_\perp) 
    \nonumber \\ 
     + \psi_K^{\uparrow,\downarrow\ast} (x,\mathbf{k}^{\prime\prime}_\perp) \psi_K^{\uparrow,\downarrow} (x,\mathbf{k}^{\prime}_\perp) \nonumber \\ 
   + \psi_K^{\downarrow,\uparrow\ast} (x,\mathbf{k}^{\prime\prime}_\perp) \psi_K^{\downarrow,\uparrow}  (x,\mathbf{k}^{\prime}_\perp) \nonumber \\ + \psi_K^{\downarrow,\downarrow\ast}  (x,\mathbf{k}^{\prime\prime}_\perp) \psi_K^{\downarrow,\downarrow}  (x,\mathbf{k}^{\prime}_\perp)\big] \, ,
\label{GPDeq}
\end{eqnarray}
where $\bfk^{\prime\prime}$ and $\bfk^{\prime}$ are the final and initial transverse momenta of quark respectively which in the symmetric plane are expressed as \cite{Puhan:2024jaw}
\begin{eqnarray}
    \bfk^{\prime\prime}&=&\bfk-(1-x)~\frac{\Dp}{2} \, , \nonumber \\
    \bfk^{\prime}&=&\bfk+(1-x)~\frac{\Dp}{2} \, .
\end{eqnarray}
At zero skewness limit, GPDs contain information about the EMFFs and PDFs of the constituent quarks. The quark EMFFs can be calculated by taking zeroth moment of the unpolarized GPDs simply by integrating over longitudinal momentum fraction as 

\begin{eqnarray}
    F^u(Q^2)=\int dx ~ H^u(x,0,t) \, ,
    \label{ffse}
\end{eqnarray}
where $-t=Q^2 (\text{GeV}^2)$. The kaon EMFF can be calculated by taking account of both flavors as 
\begin{align}
F(Q^2) &= e_u \times F^u(Q^2) + e_{\bar s} \times F^{\bar s}(Q^2) \nonumber\\
       &= e_{u} \int dx \; H^u(x,0,t) + e_{\bar s} \int dx \; H^{\bar s}(1-x,0,t),
\end{align}
where $e_{u(\bar s)}$ are the charges of $u$-quark and $\bar s$-antiquark. The kaon EMFF obey the sum rule of $F(Q^2=0)=1$ at every baryonic density.
\par The corresponding mean square charge radius of kaon can be computed as
\begin{eqnarray}
\langle r^2_K \rangle= -6 \frac{\partial F(Q^2)}{\partial Q^2}\Big|_{Q^2\rightarrow0}.
\end{eqnarray}
The charge radius of kaon with different quark flavor can be written as 
\begin{eqnarray}
  \langle r^2_K \rangle=e_u \langle r^2_u \rangle + e_{\bar s} \langle r^2_{\bar s} \rangle .
\end{eqnarray}
Using this expression, we can find the radii of different flavors. 
\par Similarly, the unpolarized valence quark PDF of pion can be obtained from unpolarized GPD at $t=0$ as 
\be
f(x)= H(x,0,0) \, .
\ee
One can also obtain the same PDF from $f_1^q(x,\bfk^2)$ TMD by integrating over transverse mometa of quark as 
\be
f(x)=\int d^2 \bfk f^q_1(x,\bfk^2).
\ee
This unpolarized PDF should obey the sum rule $\int dx f(x)=1$.

\begin{figure*}
\centering
\begin{minipage}[c]{0.98\textwidth}
(a)\includegraphics[width=7.5cm]{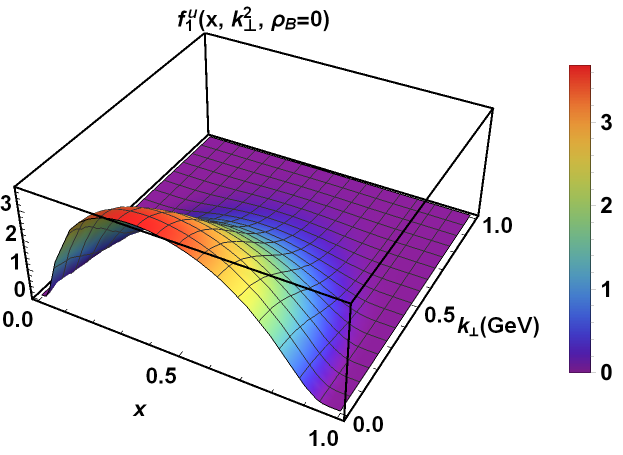}
\hspace{0.07cm}	
(b)\includegraphics[width=7.5cm]{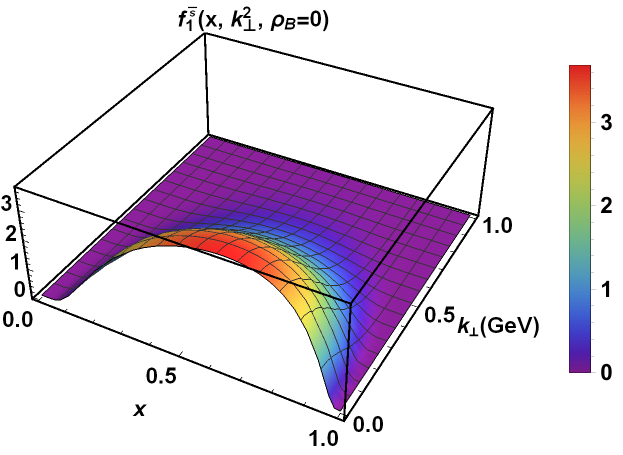}
\hspace{0.07cm}\\
(c)\includegraphics[width=7.5cm]{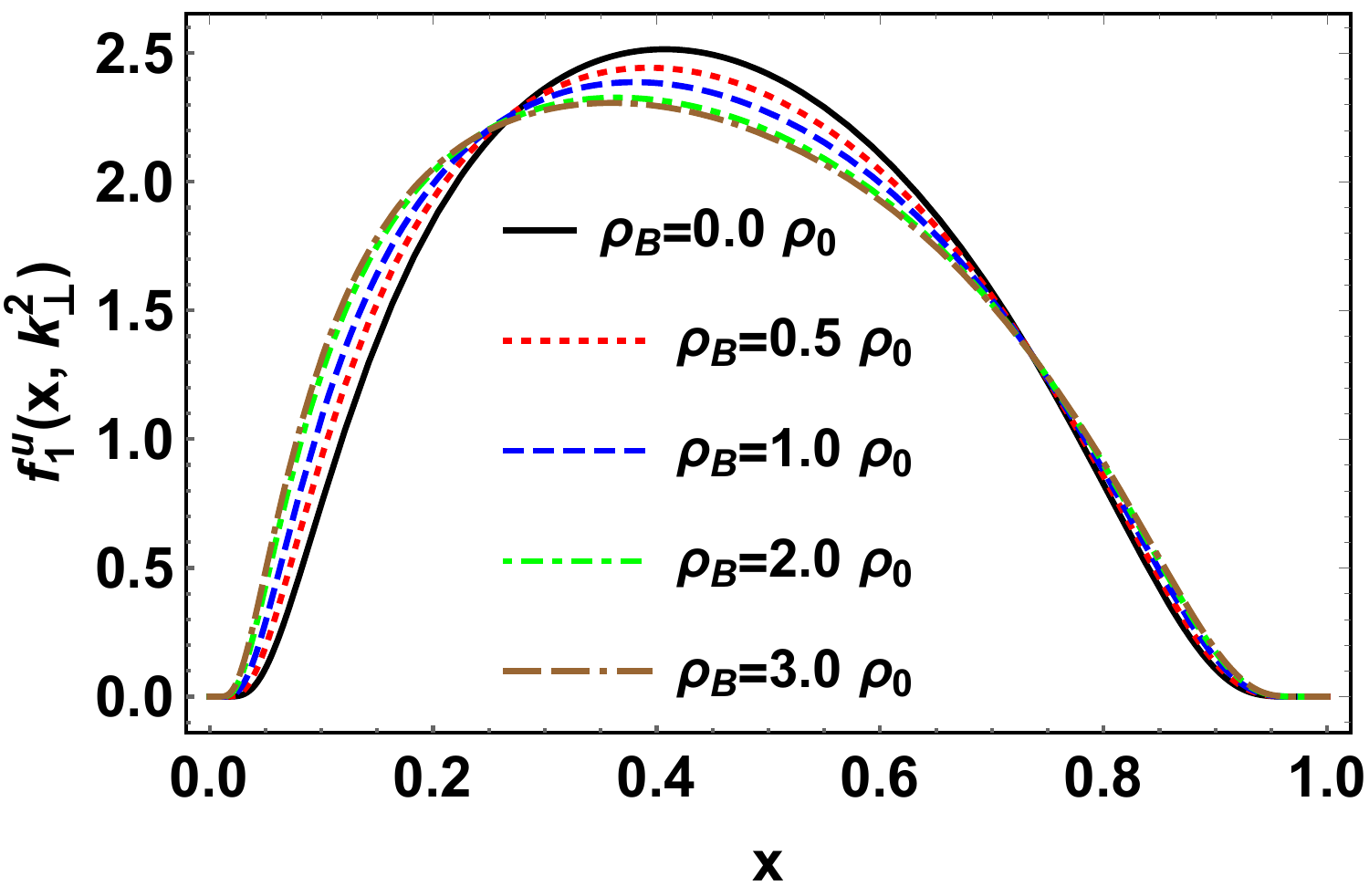}
\hspace{0.03cm} 
(d)\includegraphics[width=7.5cm]{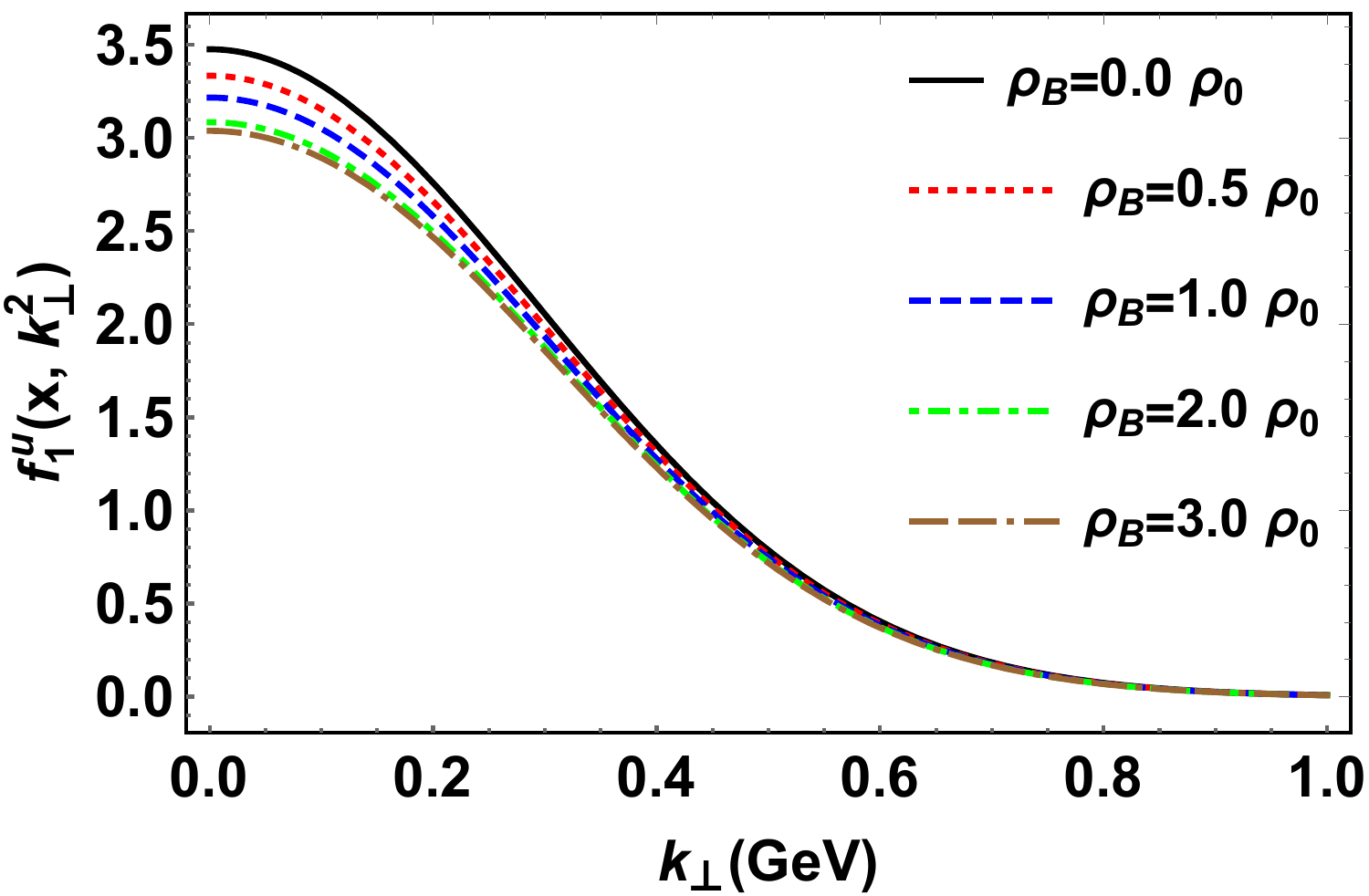}
\hspace{0.03cm}	
\end{minipage}
\caption{\label{TMDxs} (Color online) Unpolarized $f^q_1(x, \bfk^2)$ TMDs have been plotted with respect to longitudinal momentum fraction ($x$) and transverse momentum $\bfk$ for (a) $u$-quark and (b)  $\bar s$-antiquark at vacuum. The $f^u_1(x, \bfk^2)$ have been plotted at different baryonic densities with respect to (c) longitudinal momentum fraction $(x)$ at $\bfk=0.25$ GeV and (d) transverse momentum $(\bfk)$ at fixed $x=0.5$.}
\end{figure*}
\begin{figure*}
\centering
\begin{minipage}[c]{0.98\textwidth}
(a)\includegraphics[width=7.5cm]{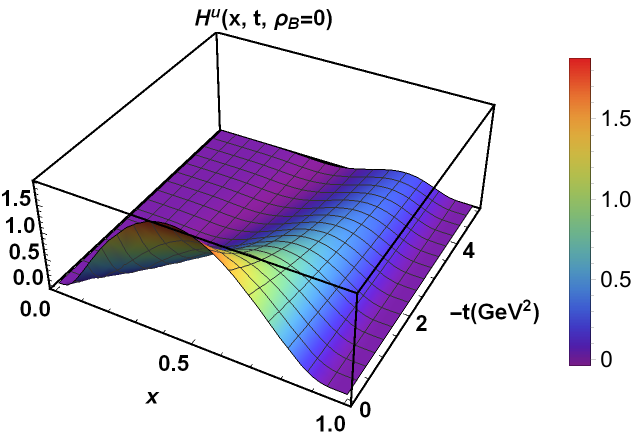}
\hspace{0.03cm}	
(b)\includegraphics[width=7.5cm]{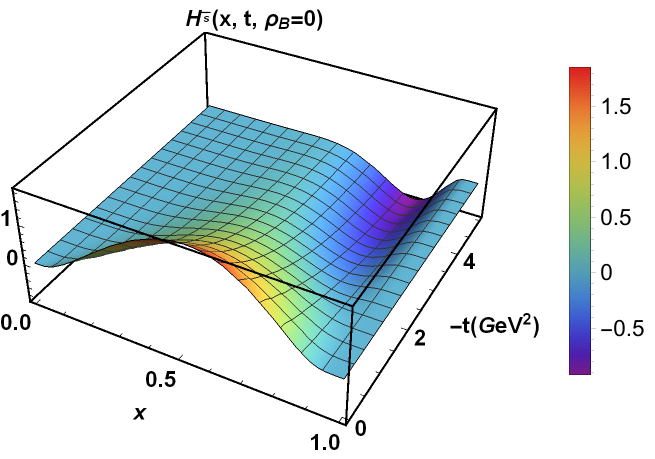}
\hspace{0.03cm}\\
(c)\includegraphics[width=7.5cm]{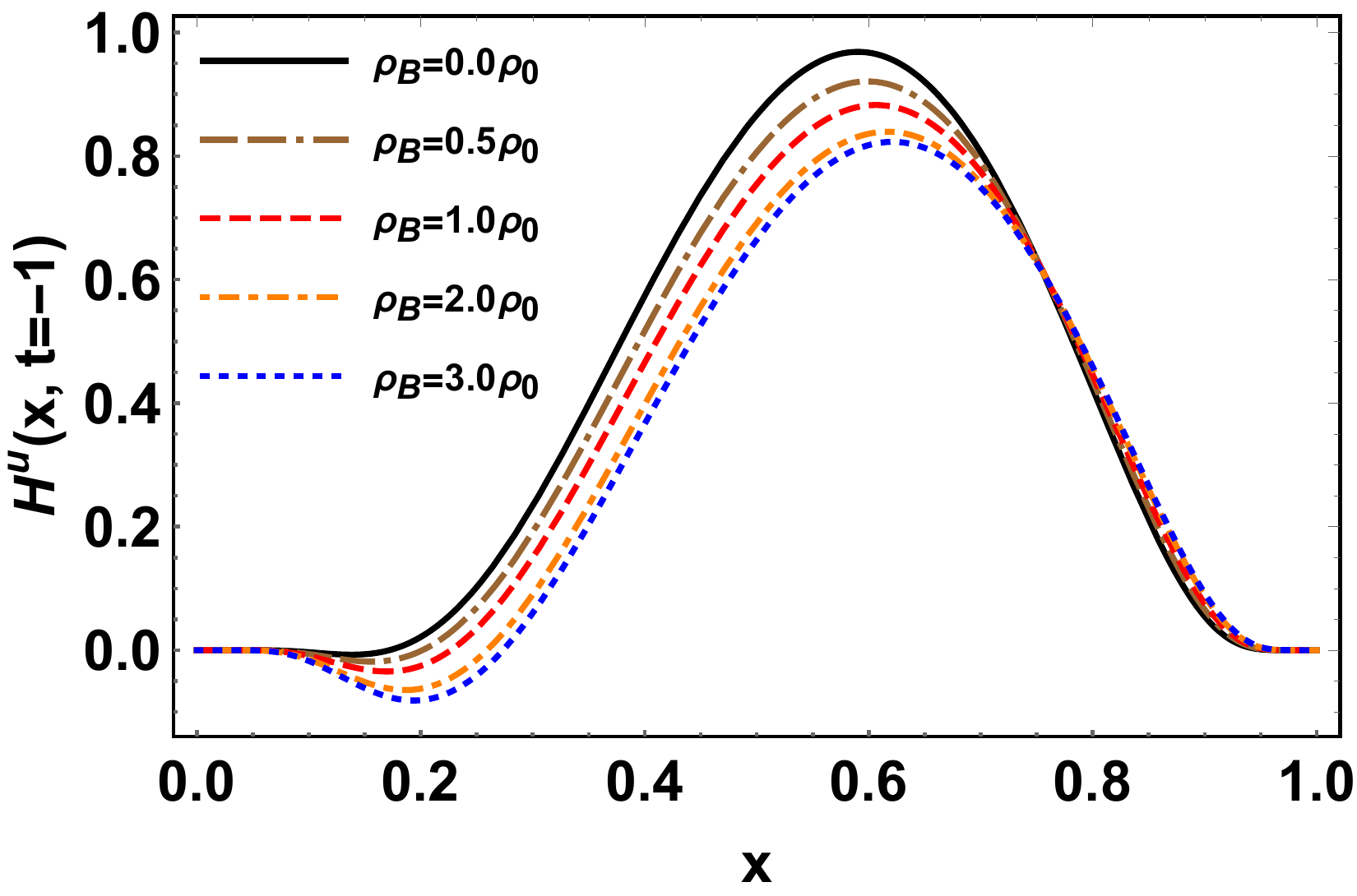}
\hspace{0.03cm} 
(d)\includegraphics[width=7.5cm]{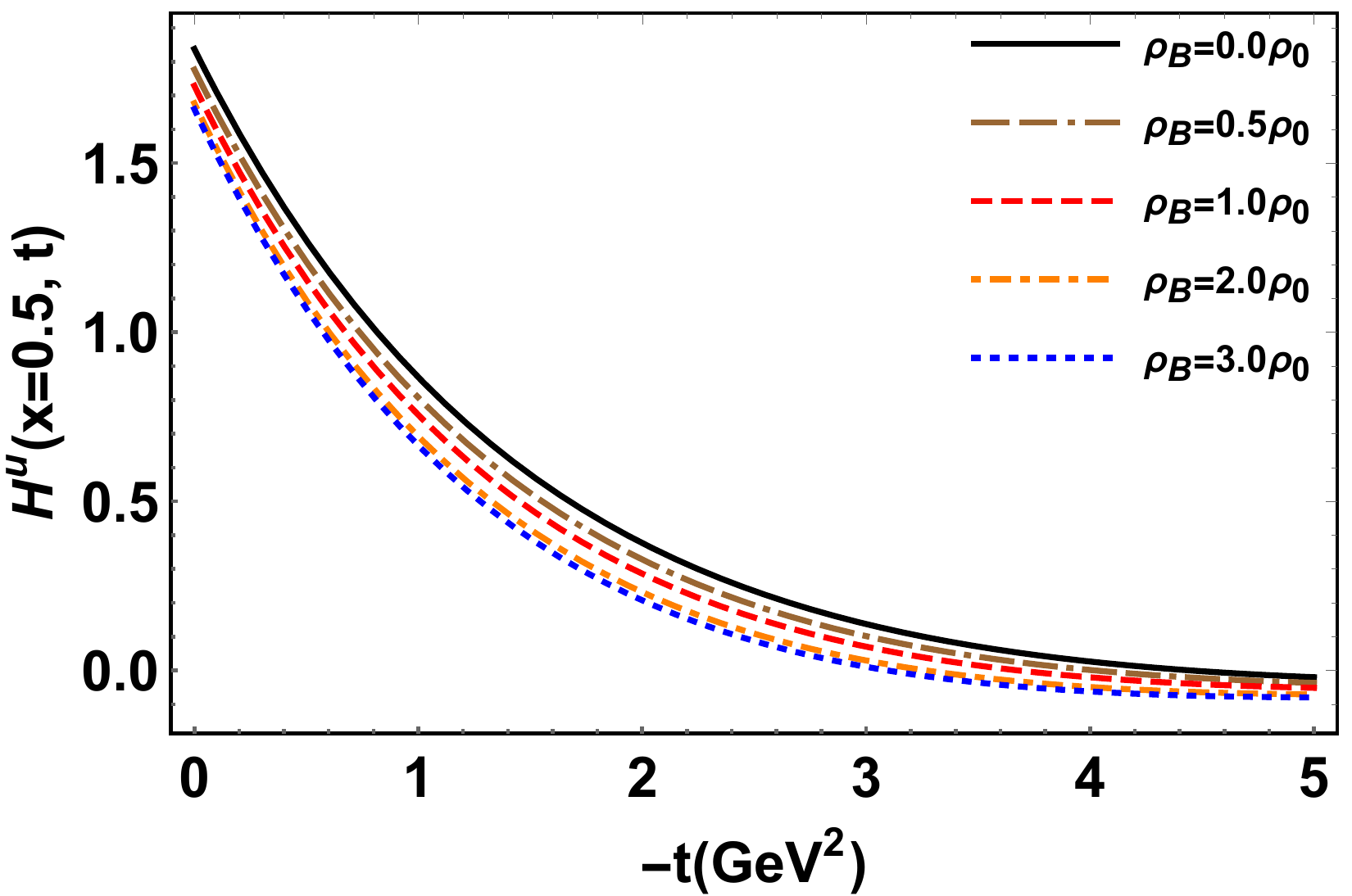}
\hspace{0.03cm}	
\end{minipage}
\caption{\label{GPDt} (Color online) Unpolarized $H^q(x, t)$ GPDs have been plotted with respect to longitudinal momentum fraction ($x$) and squared transverse momentum transfer $-t$ GeV$^2$ for (a)  $u$-quark and (b)  $\bar s$-antiquark at vacuum. The $H^u(x, t)$ have been plotted at different baryonic densities with respect to (c) longitudinal momentum fraction $(x)$ at $t= -1$ GeV$^2$ and (d) $t$ at $x=0.5$.}
\end{figure*}
\section{Results and discussion}\label{secres}
For numerical predictions of the transverse and spatial structure of kaon, we have taken the vacuum input parameter quark masses $m_{u(\bar s)}=0.256$ $(0.457)$ GeV from CQMF model and harmonic scale parameter $\beta_K=0.405$ GeV from Ref. \cite{Puhan:2023ekt}. 
The input parameters have been fitted with the kaon mass and decay constant in vacuum. 
 In Fig. \ref{TMDxs} (a) and (b), we have plotted the $f^q_1(x,\bfk^2)$ TMD for $u$-quark and $\bar s$-antiquark with respect to longitudinal momentum fraction $x$ and transverse momenta $\bfk$ in free space. Both quark TMDs vanish after $\bfk=0.8$ GeV and is smooth decreasing function with an increase in $\bfk$. Due to unequal quark masses in kaon, $u$-quark TMD shows a peak distribution towards $x\le 0.5$ whereas the heavy $\bar s$-antiquark TMD is maximum in the region $x\ge 0.5$. This indicates that heavier quarks tend to carry a greater light-front momentum fraction of the parent hadron than the lighter one. To understand the effect of nuclear medium effect, we have demonstrated the $u$-quark TMD with respect to $x$ at fixed $\bfk=0.25$ GeV in Fig. \ref{TMDxs} (c) and with respect to $\bfk$ at fixed $x=0.5$ in Fig. \ref{TMDxs} (d) at different baryonic densities upto $\rho_B= 3 \rho_0$. It is clearly illustrated that the baryonic density has considerable affect on the $u$-quark TMDs upto $x=0.7$ and $\bfk=0.5$ GeV. With an enhancement in baryonic density, the distribution decreases with $\bfk$ at fixed $x$ due to the scaling down of effective quark masses. The average transverse and longitudinal momentum fraction carried by an active quark at different baryonic densities has been presented in Table \ref{table-moment}. It is observed that the average momenta of active quark is reduced by $1.5$\% at baryonic density $3\rho_0$. After this baryonic density, there is a saturation of transverse distributions as the $\bar s$-antiquark mass decreases very slowly compared to $u$-quark mass. Similarly, the average momentum fraction carried by the active $u$-quark decreases with an increase in baryonic density, while the same increases for the $\bar s$-antiquark with an increase in baryonic density to have the total momentum fraction carried by the constituents to be unity. Overall, the nuclear medium effect is comparably less for the case of kaon than pion taking reference from our previous study on pion TMD \cite{Kaur:2024wze}.

In Fig. \ref{GPDt} (a) and (b), we have illustrated the valence quark $H^q(x,0,t)$ GPD of kaon with respect to $x$ and squared momentum transferred $t=-\Delta^2_\perp$ for $u$-quark and $\bar s$-antiquark at free space. The $u$-quark GPD shows a positive distribution upto $t=-4$ GeV$^2$, while the $\bar s$-antiquark shows both negative and positive distributions in the same region. The symmetric nuclear medium effect of $u$-quark GPD have been demonstrated at different baryonic densities with respect to $x$ at fixed $t=-1$ GeV$^2$ in Fig. \ref{GPDt} (c) and with respect to momentum transfer $-t ($GeV$^2)$ at $x=0.5$ in Fig. \ref{GPDt} (d). At a fixed value of $t$, the distribution decreases with increase in baryonic density up to $x=0.75$, indicating that the spatial distribution become more concentric and their degree of freedom increase. The difference in the in-medium distribution to free space distributions have a reasonable effect in the region $0.1\le x \le 0.7$. However, there is a reversal in the amplitude of vacuum and in-medium distributions of quark GPD around $x=0.7$. A similar kind of observation has also been observed in Ref. \cite{Puhan:2025ibn}. This indicates that with an increase of baryonic density, the momentum fraction carried by the valence quark from its parent hadron decreases and vice versa for $\bar s$-antiquark. We observed that the maximum amplitude of $H(x,t=-1)$ decreases by almost $15$\% at $\rho_B=3 \rho_0$. In the similar way, we observed that  the in-medium valence quark GPD $H^u(x,t)$ distribution decreases with an increase in baryonic density at fixed $x=0.5$ indicating the final hadron will carry less transverse momenta from initial hadron in the presence of medium. It may be  occurring due to attractive strong interaction of kaon with the surrounding nuclear medium (protons and neutrons) resulting in a decrease in the final state kaon momentum as well as valence quark momenta. 

\begin{figure*}
\centering
\begin{minipage}[c]{0.98\textwidth}
(a)\includegraphics[width=7.5cm]{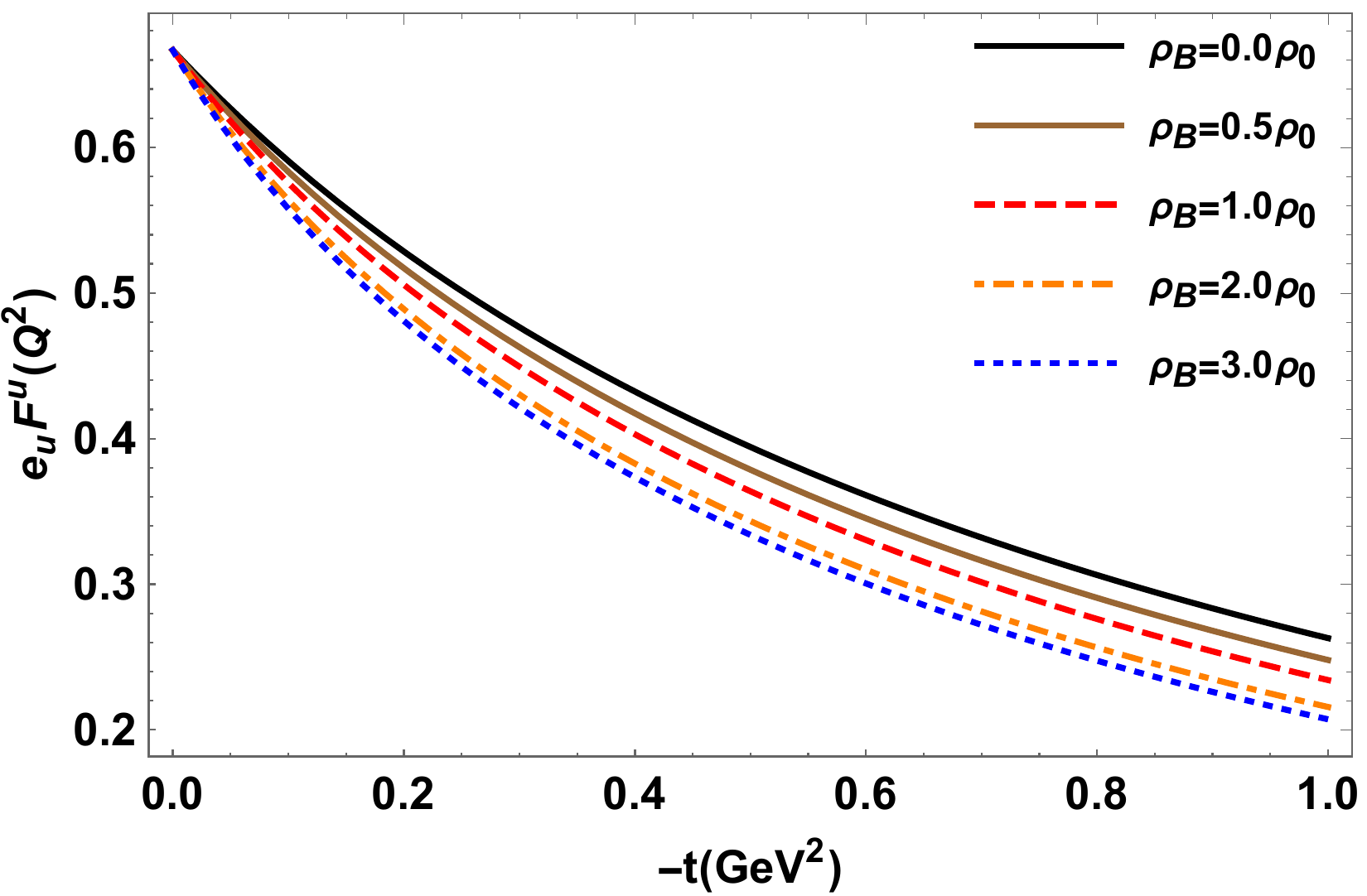}
\hspace{0.03cm}	
(b)\includegraphics[width=7.5cm]{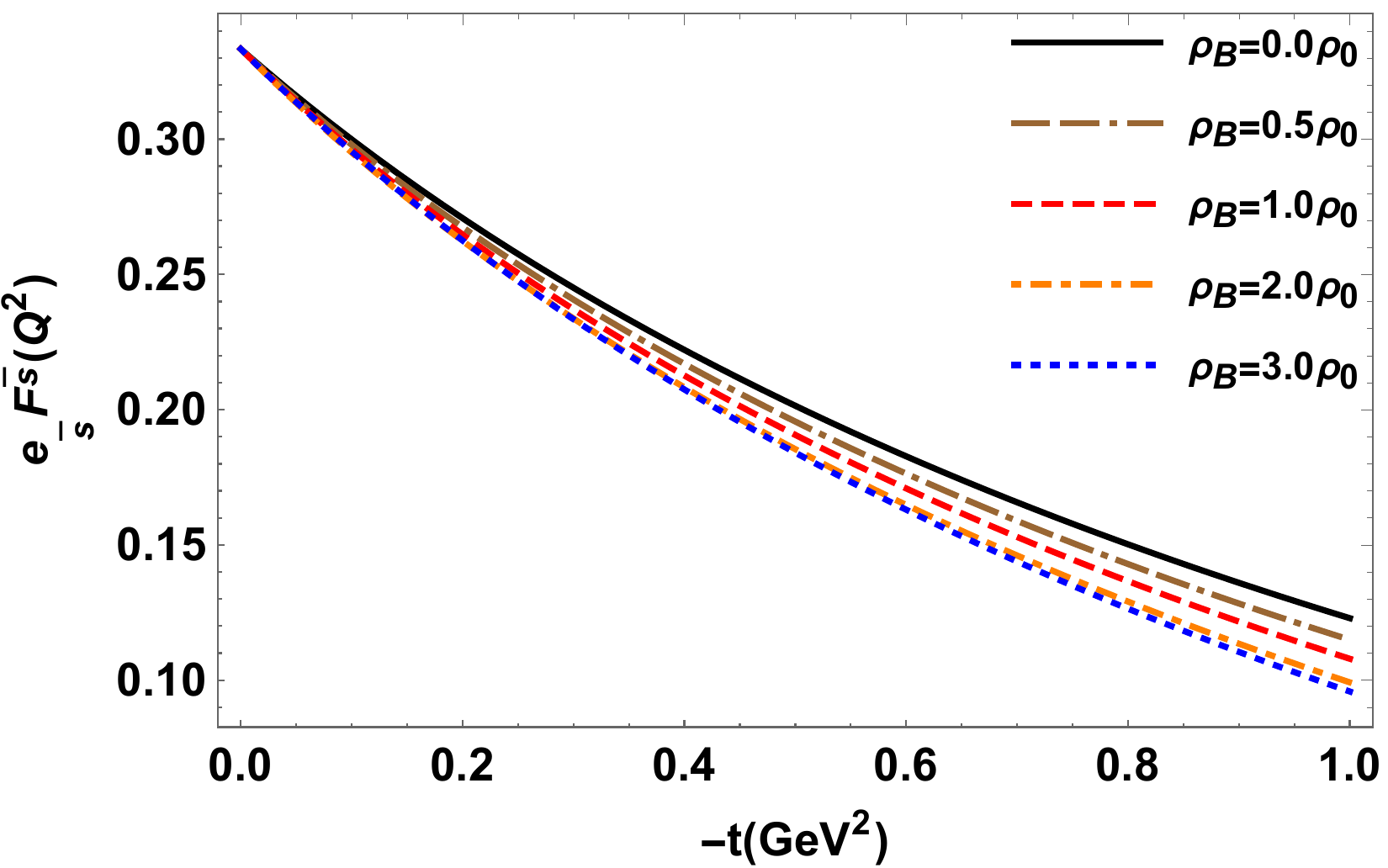}\\
\hspace{0.03cm}	
(c)\includegraphics[width=7.5cm]{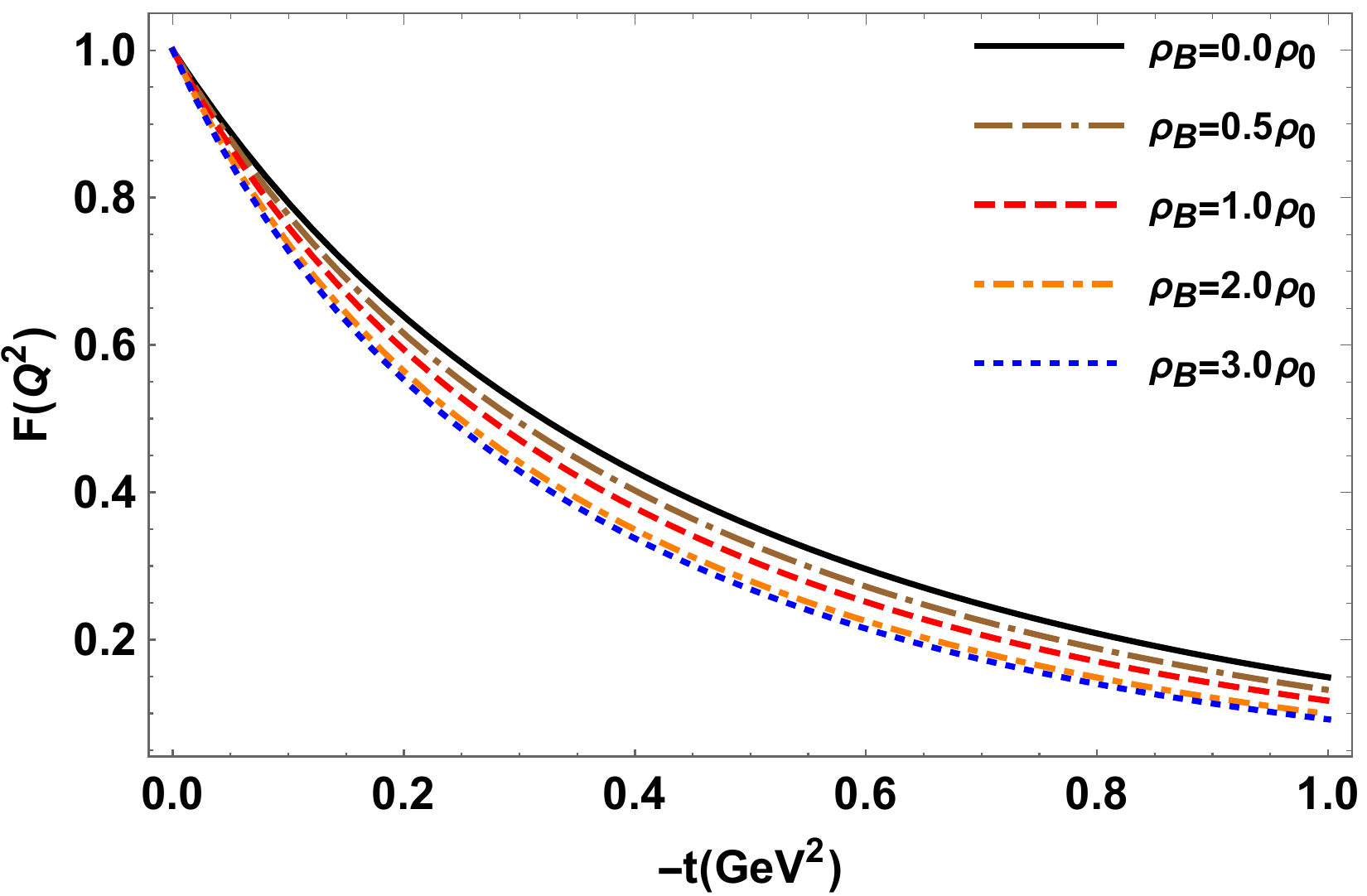}
\hspace{0.03cm}	
(d)\includegraphics[width=7.5cm]{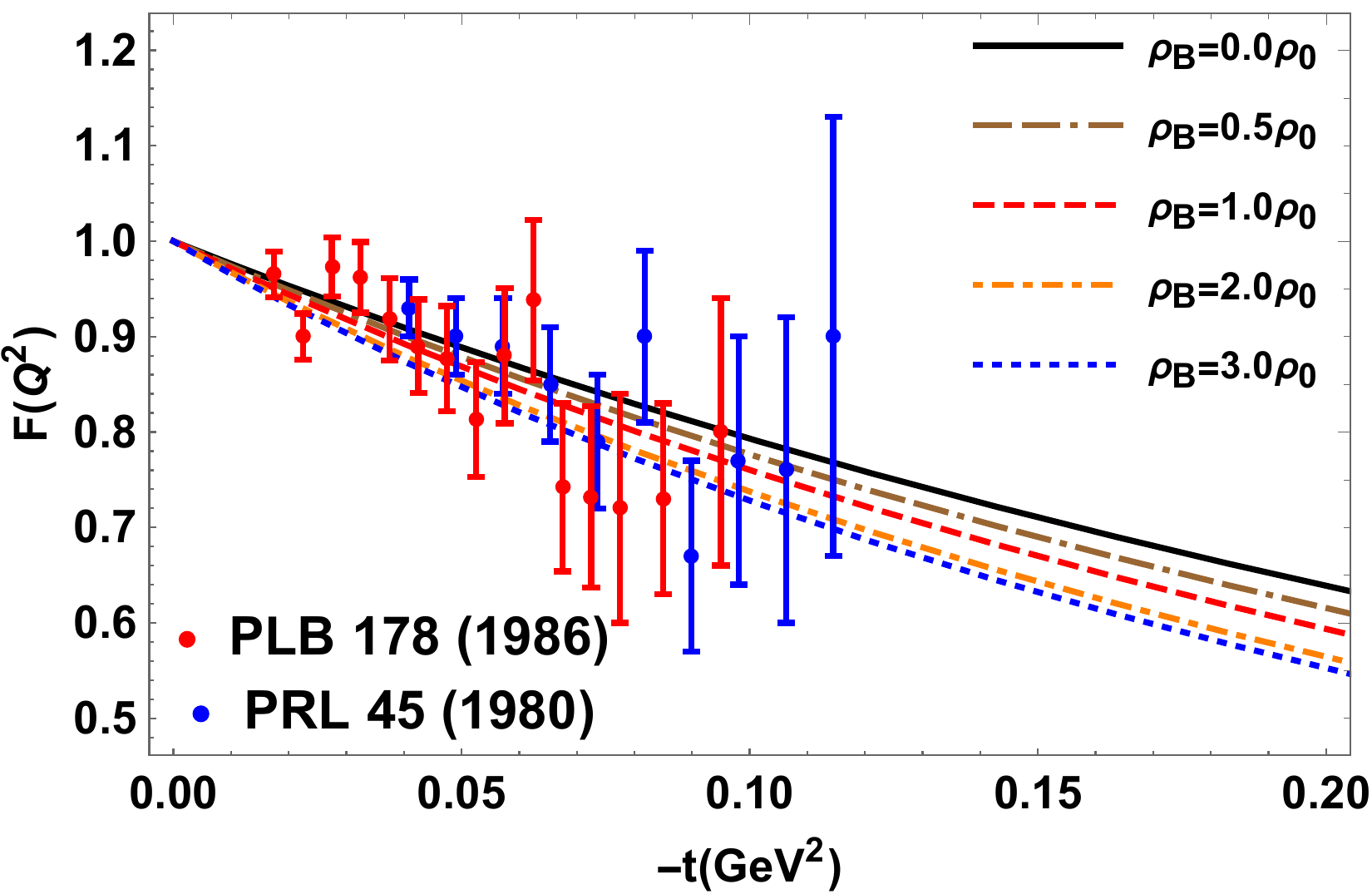}
\hspace{0.03cm}	
\end{minipage}
\caption{\label{FFwe} (Color online) Comparison of electromagnetic form factors among vacuum and in-medium distributions with different baryonic densities for (a)  $u$-quark (b)  $s$-antiquark (c) kaon and compared with experimental data \cite{Dally:1980dj,Amendolia:1986ui} in (d).}
\end{figure*}

\begin{figure}
\begin{minipage}[c]{0.98\textwidth}
\includegraphics[width=7.5cm]{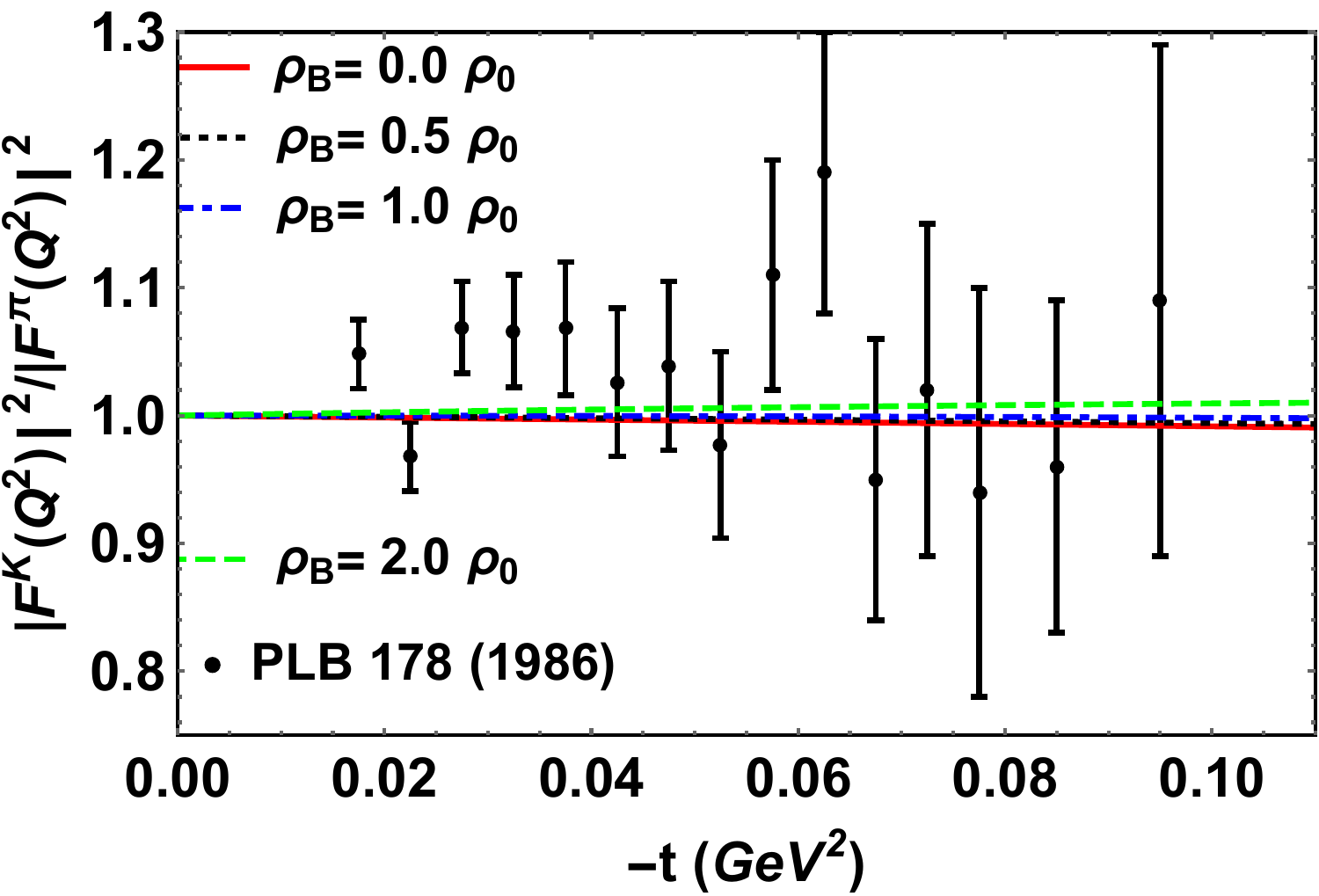}
\end{minipage}
\caption{\label{FF} (Color online) Ratio of kaon to pion EMFFs at different baryonic densities have been compared with Ref. \cite{Amendolia:1986ui}.}
\end{figure}

The EMFFs of different flavors of kaon can be calculated from their respective unpolarized GPDs by using Eq. (\ref{ffse}) and have been plotted with respect to $-t$ (GeV$^2$) in Fig. \ref{FFwe} (a) and (b) for $u$-quark and $\bar s$-antiquark with their charges respectively. The $u$-quark EMMFs show a comparably higher decrease in nuclear medium compared to $\bar s$-antiquark. These flavor EMFFs decrease smoothly as the medium become denser resulting in a decrease in their charge distributions. Similar observations have also been observed in Ref. \cite{Yabusaki:2023zin}. At $Q^2=0$, the absolute value of $|e_uF^u(Q^2)|$ and $|e_{\bar s}F^ 
{\bar s}(Q^2)|$ are found to be $0.668$ and $0.332$ at free space as well as in-medium respectively. The total contribution coming from both the flavors is $1$ obeying the EMFFs sum rule $F(Q^2=0)=1$. The total kaon EMFFs have been plotted at different baryonic densities in Fig. \ref{FFwe}(c) and have been compared with available experimental data in Fig. \ref{FFwe}(d) \cite{Dally:1980dj,Amendolia:1986ui}. The free space and in-medium kaon EMFFs are found to be in good agreement with the experimental data. As the medium becomes more and more dense, the interaction of kaon with the nucleons increases, resulting in the decrement of EMFFs distributions with $-t$.
We have also plotted the kaon to pion EMFFs ratio in Fig. \ref{FF}. The pion in-medium results have been taken from our previous study \cite{Puhan:2025ibn}. The ratio is found to be in good agreement with only available data \cite{Amendolia:1986ui} up to  baryonic density $2 \rho_0$. The ratio increases with an increase in baryonic density as the $\bar s$-antiquark interaction is less compared to $u$-quark in medium. It is important to mention here that this is an unique phenomena observed in this study. As there is no experimental data and theoretical predictions available to validate this kind of behavior, we will need further investigation on this.

We have also calculated the charge mean square radius of kaon along with its flavors in the units of fm$^2$ and cited in the Table \ref{table-moment}. The free space charge radius of kaon is found to be $0.259$ fm$^2$ compared to the experimental result of $0.34$ fm$^2$ \cite{Amendolia:1986ui} with a deviation of $23$\%. With the increase in baryonic density  the radius increases, and this deviation is found to be $6$\%  and $14$\% at $2\rho_0$ and $3\rho_0$ baryonic density, respectively.
 In the Table we also compare the charge radius of kaon with the lattice simulation result \cite{Aoki:2015pba}. Similarly, the charge mean square radius of $u$-quark is found to increase faster than the $\bar s$-antiquark. This occurs as $u$-quark mass decreases more rapidly than $\bar s$-antiquark resulting in the increment of spatial distribution and a wider region to distribute its charge. The charge mean square radii of $u$-quark and $\bar s$-antiquark are found to be increased by $18$\% and $2.7$\% at $3 \rho_0$ from free space. We have also compared our results with available theoretical predictions \cite{Arifi:2024tix,Hutauruk:2019was,Gao:2017mmp} in Table \ref{table-moment} and a similar kind of behavior is followed. The absolute value of the charge mean square radii difference of pion to kaon has also been explored in this work and has been presented for different baryonic densities in Table \ref{table-moment}. From this study, we can conclude that with increase in baryonic density, the quark-antiquark mass decreases resulting in the increment of larger space for both flavors. This results in a decrease in the binding energy of quark-antiquark which implies that kaon is less bound compared to free space, due to which the kaon radius increases.

\section{Summary and conclusions}\label{secsum} In this present work, we have demonstrated the effect of symmetric nuclear matter on the transverse and spatial structure of kaon using the light-cone quark model (LCQM) and chiral SU($3$) quark mean field model (CQMF). The transverse structure of the valence quark has been studied using transverse momentum parton distribution functions (TMDs) and the spatial structure using the generalized parton distribution functions (GPDs). Both valence quark TMDs and GPDs have been calculated by solving the quark quark correlator using LCQM. The in-medium properties have been explored by modifying the input variables in CQMF. We found that the $u$-quark mass decreases rapidly as compared to $\bar s$-antiquark with baryonic density. The in-medium valence quark is found to be carry less longitudinal momentum fraction ($x$) and transverse momenta ($\bfk$) from its parent hadron compared to free space. We found that the average momenta of $u$-quark is reduced by $5$\% in the presence of nuclear medium at $\rho_B=3 \rho_0$ than the free space. We have also found that the transferred transverse momentum between final and initial hadron decreases with increase in baryonic density and the distribution saturates after $\rho_B=3 \rho_0$. Similar kind of observations have been obtained for the case of pion in our previous study. For a better understanding of nuclear medium effect, we have calculated the electromagnetic form factors (EMFFs) and charge mean square radii of kaon along with its quark flavors. These results are found to be in good agreement with available experimental data. The effect of medium on EMFFs and mean square radii is found to be high in the case of $u$-quark than the $\bar s$-antiquark. This indicates that the spatial distribution of light quark masses are highly affected by medium as compared to the heavier quark mass. The mean square radius of kaon is found to decrease with an increase in baryonic density allowing quarks to move in a wide region due to the decrease in binding energy of quark antiquark inside the kaon. The free space kaon to pion EMFFs ratio is also found to be in good agreement with the experimental data. This ratio is found to be increasing in the presence of nuclear medium. Overall, there is considerable effect of nuclear medium on the valence quark distribution of kaon. The observed free space and in-medium properties of kaon will provide fruitful insights for the future measurements of kaon in J-PARC in Japan \cite{Aoki:2021cqa} and EIC in US \cite{AbdulKhalek:2021gbh}.

\begin{table*}
\begin{tabular}{|c|c|c|c|c|c|c|c|c|c|c|c|}
\hline 
Baryonic & \multicolumn{7}{c|}{LCQM } & \multicolumn{3}{c|}{NJL model}& \multicolumn{1}{c|}{LFQM}\\
 density $(\rho_B/\rho_0)$ & \multicolumn{7}{c|}{(This work)} & \multicolumn{3}{c|}{\cite{Hutauruk:2019was}}& \multicolumn{1}{c|}{\cite{Arifi:2024tix}} \\
\cline{2-12}
 & $\langle \bfk^u \rangle$ & $\langle x^u \rangle$& $\langle x^{\bar s} \rangle$& $\langle r^2_u \rangle$ & $\langle r^2_{\bar s} \rangle$ & $\langle r^2_K \rangle$  & $\langle r^2_\pi \rangle -\langle r^2_K \rangle$ & $\langle r^2_u \rangle$  & $\langle r^2_{\bar s} \rangle$ & $\langle r^2_K \rangle$  & $\langle r^2_K \rangle$  \\
\hline
$0$ & 0.333 & 0.455&0.545&0.297 & 0.182 & 0.259 & 0.015&0.423 & 0.194 & 0.348 & 0.36\\
$0.5$ & 0.331&0.450& 0.550 & 0.332 & 0.189 & 0.284& 0.022 & 0.476 & 0.194 & 0.384 & 0.485\\
$1$ & 0.330 &0.446& 0.554 & 0.371 & 0.195 & 0.313& 0.029 & O.624 & 0.194 & 0.476 & 0.697\\
$2$ & 0.328& 0.442 &0.558 & 0.440 & 0.204 & 0.361& 0.018 &-&-&-&-\\
$3$ & 0.328& 0.440 & 0.560 & 0.480 & 0.209 & 0.389 & - & -& -& -&-\\
Exp. \cite{Amendolia:1986ui}&-&- &  - & - & - &0.34$\pm0.05$ &0.10 $\pm$ 0.45 & -& -& -&-\\
Lat. \cite{Aoki:2015pba} &-&- &  -  & - & - & $0.38\pm 0.012$& -& -& -&-&-\\
Ref. \cite{Gao:2017mmp} &-&- &  -  & 0.384 & 0.184 & -& -& -& -&-&-\\
\hline
\end{tabular}
\caption{The average transverse momenta of $u$-quark, the average momentum fraction carried by different flavors $\langle x^{u(\bar s)}\rangle$, the charge radii of different flavors and kaon have been presented at different baryonic density and have been compared with other model predictions \cite{Hutauruk:2019was,Arifi:2024tix} along with experimental data \cite{Amendolia:1986ui} and lattice simulation data \cite{Aoki:2015pba} at vacuum. The mean square charge radius are presented in the unit of fm$^2$.}
\label{table-moment}
\end{table*}



\section*{Acknowledgement}
H. D.  would like to thank  the Science and Engineering Research Board,  Anusandhan-National Research Foundation,  Government of India under the scheme SERB-POWER Fellowship (Ref No.  SPF/2023/000116) for financial support. 
\appendix



\end{document}